\documentclass[12pt]{article}

\usepackage{amssymb}
\usepackage{amsmath}
\usepackage[mathscr]{euscript}

\input{epsf}

\newcommand{\beq}{\begin{eqnarray}}
\newcommand{\eeq}{\end{eqnarray}}
\newcommand{\half}{\frac{1}{2}}
\newcommand{\thalf}{{\textstyle{1 \over 2}}}
\newcommand{\tthhalf}{{\textstyle{3 \over 2}}}
\newcommand{\del}{\partial}

\newcommand{\ket}{\rangle}
\newcommand{\bra}{\langle}

\newcommand{\fpi}{f_{\pi}}

\newcommand{\dslash}{{\del \hspace{-6pt}/}}

\begin{document}

\begin{center}

{\large Determination of the axial coupling constant $g_{A}$ in the 
 linear representations of chiral symmetry}

\vspace*{0.5cm}

{Atsushi Hosaka\\
Research Center for Nuclear Physics (RCNP), Osaka Univ. 
Ibaraki 567-0047 Japan~\footnote{
E-mail: hosaka@rcnp.osaka-u.ac.jp}}\\[0.2cm]
{Daisuke Jido\\
Consejo Superior de
 Investigaciones Cient\'{\i}ficas, Universitat de Valencia, IFIC,
 Institutos de Investigaci\'{o}n de Peterna, Aptdo.
 Correos 2085, 46071, Valencia, Spain}\\[0.2cm]
{Makoto Oka\\
Department of Physics, 
Tokyo Institute of Technology,
 Meguro, Tokyo 152-8551 Japan}

\end{center}

%


\abstract{
If a baryon field belongs to a certain linear representation of chiral 
symmetry of $SU(2) \otimes SU(2)$, the axial coupling constant $g_{A}$ 
can be determined algebraically from the commutation relations derived 
from the superconvergence property of pion-nucleon scattering amplitudes.  
This establishes an algebraic explanation for the values of $g_{A}$ 
of such as the non-relativistic quark model, large-$N_{c}$ limit and 
the mirror assignment for two chiral partner nucleons. 
For the mirror assignment, the axial charges of the positive and negative 
parity nucleons have opposite signs.  
Experiments of eta and pion productions 
are proposed in which the sign difference of the axial 
charges can be observed.   
}

\section{Introduction}

The axial coupling constant $g_{A}$ of the nucleon is one of 
fundamental constants of the nucleon; 
experimentally, it is $g_{A} = 1.25$.  
Sometimes, the fact that $g_{A}$ is close to unity is considered 
as an evidence of partially conserved axial vector current (PCAC), and 
hence it would become one in the limit that the axial current is 
conserved.  
This, however, is not correct, and $g_{A}$ 
takes any number when chiral symmetry is broken 
spontaneously~\cite{weinberg1996}. 
This fact is manifest in the non-linear sigma model.  
Even in the linear sigma model, $g_{A}$ can be arbitrary if higher 
derivative terms are added.  

Theoretically, $g_{A}$ is related to the generators of the chiral 
group:
\beq
\left[ Q_{V}^a, Q_{V}^b \right] = i \epsilon_{abc} Q_{V}^c \, , \; \; 
\left[ Q_{V}^a, Q_{A}^b \right] = i \epsilon_{abc} Q_{A}^c \, , \; \; 
\left[ Q_{A}^a, Q_{A}^b \right] = i \epsilon_{abc} Q_{A}^c \, .
\label{Qalbebra} 
\eeq
If chiral symmetry is not spontaneously broken, these commutation 
relations may be used to determine the value of $g_{A}$, the nucleon
matrix element of the axial charge operator $Q_{A}^a$.  
When, however, chiral symmetry is spontaneously broken, 
the operators $Q_{A}^a$ are not well defined. 
In his pioneering work, Weinberg used the pion-nucleon matrix 
elements rather than the axial charges $Q_{A}^a$ to compute $g_{A}$ by 
using Goldberger-Treiman relation and commutation relations among the 
pion-nucleon coupling matrices $X^a$~\cite{weinberg1969}.   
These commutation relations are derived from a super convergence 
property of pion-nucleon scattering amplitudes; it is a consistency 
condition between the low momentum expansion and the asymptotic 
behavior of the amplitudes.  
In the dispersion theory, Aldler and Weisberger derived a sum rule 
from the commutation relations~\cite{adler1965,weisberger1966}.   
Weinberg showed that when continuum intermediate states were saturated 
by narrow one particle states, the dispersion relations 
reduce to a set of algebraic equations which are in some cases solved 
to provide the value of $g_{A}$.  

From a group theoretical point of view, a closed algebra determines 
the values of charges.  
A rather trivial example is the isospin charge of $SU(2)$.  
Similarly the axial part of the chiral group 
(a coset $SU(2) \times SU(2)/SU(2)$) when combined with the isospin 
part determines the axial charge of 
a linear representation of the full chiral 
group $SU(2) \times SU(2)$.   
In this paper we discuss several examples where $g_{A}$ 
can be determined by the commutation relations, including the cases 
corresponding to the non-relativistic quark model, large-$N_{c}$
limit and the mirror assignment for the two nucleons of chiral 
partners.  

The mirror representation of the chiral group for the nucleon is 
particularly interesting, since it provides a possibility where 
positive and negative parity nucleons belong to the same 
chiral multiplet, showing characteristic behaviors toward chiral symmetry 
restoration~\cite{jido2000}.  
In the latter part of this paper, we propose experimental method in 
which we will be able to observe the mirror nucleons in pion and eta 
productions from the nucleon~\cite{jido2001}.

\section{Algebraic determination of axial charges}

The use of the algebraic method was considered 
by Weinberg long ago~\cite{weinberg1969}.  
He computed pion-nucleon coupling matrix elements for various 
linear representations of the chiral group.  
This method was later used in explaining the reason that the 
axial charge 
and magnetic moments of the constituent quarks takes the bare 
values~\cite{weinberg1990}.   

It was shown that commutation relations among the pion-nucleon 
coupling matrices $X_{a}$ and isospin charges $T^a$,
\beq
\label{commX}
\left[ X^a , \, X^b \right] 
= i \epsilon_{abc} T^c \, , \; \; \; 
\left[ T^a , \, T^b\right] = i \epsilon_{abc} T^c \, , \; \; \; 
\left[ T^a , \, X^b\right] = i \epsilon_{abc} X^c\, .
\label{commTX}
\eeq
were derived by considering the asymptotic behavior of the 
pion-nucleon forward scattering 
amplitudes when they are computed from a low energy effective lagrangian.  
Here the matrices $X^a$ and $T^a$ are 
related to the matrix elements of the axial vector and vector currents, 
\beq
\label{NAN}
\bra N_{\beta} \lambda^\prime |
\int d^3 x \, \tilde A_{3}^a(x) | N_{\alpha} \lambda \ket 
&\equiv& 
( X^a )_{\beta\lambda^\prime, \alpha\lambda} \, , \\
\label{NVN}
\bra N_{\beta} \lambda^\prime | 
\int d^3 x \, V_{0}^a(x) | N_{\alpha} \lambda \ket 
&\equiv&
\left( T^a \right)_{\beta\lambda^\prime, \alpha\lambda} \, .  
\eeq
Here $\alpha$ and $\beta$ are isospin indices of the nucleon, and 
$\lambda$ and $\lambda^\prime$ are the helicities.  
When the momentum direction is taken along the $z$-axis, 
the matrices $X^a$ and $T^a$ are related to the axial charge 
$g_{A}$ and isospin charge $g_{V}=1$ by 
$
( X^a )_{\beta\lambda^\prime, \alpha\lambda} =
g_{A} ( \tau^a / 2 )_{\beta \alpha} 
(\sigma_{3})_{\lambda^\prime \lambda}$, 
$( T^a )_{\beta \alpha} 
=
g_{V} (\tau^a/2)_{\beta \alpha} \delta_{\lambda^\prime \lambda}$.  
The currents appear in the low energy effective 
lagrangian~\cite{weinberg1996}
\beq
\label{Lint}
L_{int} 
=
\frac{2i}{f_{\pi}^2} V_{\mu}^a \epsilon_{abc} \pi_{b} \del^\mu \pi_{c} 
+ 
\frac{1}{f_{\pi}^2} \tilde A_{\mu} ^a \del^\mu \pi_{a} \, , 
\eeq
where 
$\tilde A_{\mu}^a = A_{\mu}^a - (-f_{\pi} \del_{\mu} \pi^a)$.  
In (\ref{Lint}), the pion-nucleon coupling constant $g$ is replaced 
by $g_{A}M/\fpi$ through the Goldberger-Treiman relation.  

The scattering amplitudes computed by using the lagrangian 
(\ref{Lint}) are small momentum expansion around $p = 0$, reproducing 
the low energy behavior expected from the low energy theorems.  
However, the large momentum behavior of the amplitudes
is not consistent with the lower bound of unitarity.  
The commutation relations (\ref{commX}) are introduced in order to 
reproduce the correct asymptotic behavior of the 
amplitudes~\cite{weinberg1969}.   

\begin{figure}[tbp]
    \centering
    \footnotesize
    \epsfxsize = 11cm
    \epsfbox{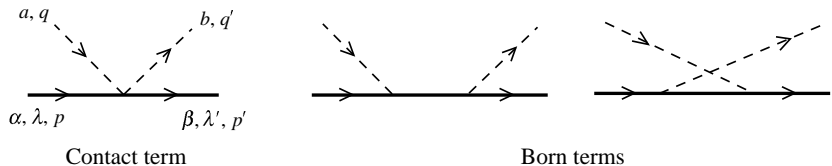}
    \begin{minipage}{11cm}
	\centering
	\caption{ \small
 Contact and Born diagrams for the pion-nucleon scatterings. 
	\label{pinscatt}}
    \end{minipage}
\end{figure}

Now using the commutation relations, we  can determine the charges, 
the matrix elements of $X^a$ and $T^a$.  
The isospin charge is trivial; it is normalized, $g_{V}=1$.  
Now taking the 
nucleon  matrix elements, one can 
compute the axial charges (pion-nucleon couplings) $X^a$.  
For example, we consider a linear representation 
$(\thalf, 1) \oplus  (1, \thalf)$.  
The numbers in the parentheses represent isospin values of 
$SU(2) \otimes SU(2)$.  
Then
the first term of the direct sum, $(\thalf, 1)$, 
is the representation for the 
chirality plus component $\psi_{+} = \thalf (1+\gamma_{5}) \psi$, 
and the second term, $(1, \thalf)$, for the chirality minus component 
$\psi_{-} = \thalf (1-\gamma_{5}) \psi$.  
The chiral representation $(\thalf, 1)$ contains terms of isospin 1/2 
(nucleon) and 3/2  (delta)  as diagonal combinations of 1/2  and 1.  

To be specific, 
let us consider matrix elements of the commutation relation
\beq
\left[ X^+, \; X^- \right] = -T^3 \
\eeq
between
$| N \ket = |IM\ket = |1/2\; 1/2\ket$ and 
$| \Delta \ket = |IM\ket = |3/2\; 1/2\ket$.  
Here 
$X^{\pm} = \mp 1/\sqrt{2} (X_{1} \pm i X_{2})$.   
Writing the reduced matrix elements as
$\bra N || X || N \ket \equiv X_{N}$, 
$\bra N || X || \Delta \ket \equiv X_{N\Delta}$ and 
$\bra \Delta || X || \Delta \ket \equiv X_{\Delta}$, 
we obtain the following three coupled equations:
\beq
& & 
- \frac{1}{3} X_{N}^2 + \frac{1}{6} X_{N\Delta}^2 = - \half \, , 
\nonumber \\
& & \frac{1}{30} X_{\Delta}^2 + \frac{1}{12} X_{N\Delta}^2 = \half \, , 
\label{coupleX} \\
& &
2X_{N} X_{N\Delta} - \sqrt{10} X_{\Delta} X_{N\Delta} = 0 \, .
\nonumber
\eeq
Solving these coupled equations for $X_{N\Delta} \neq 0$, 
we find 
\beq
|X_{N}| = \frac{5}{\sqrt{6}} \, , \; \; \; 
|X_{\Delta}| = \sqrt{\frac{5}{3}} \, , \; \; \; 
|X_{N\Delta}| = \frac{4}{\sqrt{3}} \, .
\eeq
These results lead to the nucleon axial charge
$|g_{A}^N| = 5/3$, 
which is the value of the non-relativistic quark model.  
This agreement is not accidental, since
in the quark model, the nucleon and 
delta can be described in the same basis of three quarks, as 
corresponding to the chiral representation 
$(1,\,  \thalf) \oplus (\thalf ,\,  1)$.  
If $X_{N\Delta} = 0$, 
there is no coupling between the nucleon and delta, where they 
belong to separate representations; 
$N \sim (\thalf, 0) \oplus (0, \thalf)$ and 
$\Delta \sim (\tthhalf, 0) \oplus (0, \tthhalf)$.  
In this case the nucleon axial charge reduces simply to unity, 
$g_{A}= 1$. 
This explains the result of the linear sigma model.  

We can extend this analysis to various cases (representations).  
Here we show two examples; one is the large-$N_c$ limit and the other 
is the mirror assignment for parity doublet nucleons.  
The large-$N_{c}$ nucleons are represented by the representation
$((N_{c}+1)/4, (N_{c}-1)/4) \oplus ((N_{c}-1)/4, 
(N_{c}+1)/4)$~\cite{beane}.  
Obviously, this representation contains the isospin states of 
$ 1/2 \le I \le N_{c}/2$.  
The relevant equations corresponding to (\ref{coupleX}) are recursion 
equations from $I = 1/2$ to $N_{c}/2$.  
The solution gives the nucleon $g_{A} = (N_{c} +2)/3$, which is the 
result known in the large-$N_{c}$ quark model~\cite{manohar}.  
In the mirror assignment, we consider two nucleons of opposite parity 
(parity doublet).  
Then, the assignment of the chiral group $(I_{1}, I_{2})$ for the 
chirality plus and minus components is interchanged for the two 
nucleons;
\beq
|1\ket : \; 
\psi_{+} \sim (\thalf, 0)\, , \;
\psi_{-} \sim (0, \thalf)\, , \; \; \; 
|2\ket : \; 
\psi_{+} \sim (0, \thalf)\, , \; 
\psi_{-} \sim (\thalf, 0)\, . 
\label{mirror}
\eeq
Obviously, the absolute values of 
the axial charges of the two nucleons are unity 
but with opposite signs.  
It is also possible to assign the same chiral representation to the 
two nucleons.  
This assignment was called the naive assignment.  
In general, the two representations of (\ref{mirror}) can mix in 
physical nucleons;  
the chiral eigenstates and mass eigenstates differ.  
Introducing a mixing angle $\theta$, the axial charges of the two 
physical nucleons (mass eigenstates) are given in the matrix form
(see Eq. (\ref{mirq5com})
\beq
\label{gAmirror}
g_{A} = \left(
\begin{array}{c c}
    \cos 2\theta & -\sin 2\theta \\
    -\sin 2\theta & - \cos 2 \theta
\end{array}
\right) \, .
\eeq

\section{Experimental observation of the mirror assignment}

Among the above examples, the mirror (as well as the naive) 
representations were not considered much 
before.  
The possibility that the axial charge of the negative parity nucleon 
can be opposite to the nucleon axial charge was first pointed out by 
Lee and realized in the form of the linear sigma model by DeTar and 
Kunihiro.  
Weinberg also pointed out the two possibilities 
$g_{A} = \pm 1$~\cite{weinberg1990}.  
At the composite level, the nucleon axial charge should be derived 
from the underlying theory of QCD.  
So far, we do not know reliable methods to do so.  
In this section, therefore, we first 
present phenomenological properties of the 
two chiral assignments, the naive and mirror, using the linear sigma 
model~\cite{jido2000}.  
We then propose an experimental method to observe the two  
assignments~\cite{jido2001}.

Let us consider linear sigma models based on the naive and mirror 
assignments.  
To do this, 
meson fields are introduced as components of the representation 
$(1/2, 1/2)$ of the chiral group, 
which are subject to the transformation rule:
$
\sigma + i \vec \tau \cdot \vec \pi 
\to
g_{L} (\sigma + i \vec \tau \cdot \vec \pi) g_{R}^\dagger \, .
$

In the naive assignment, the chiral invariant 
lagrangian up to order (mass)$^4$ 
is given by: 
\beq
L_{\rm{naive}} 
& = & 
\bar{N_1} i \dslash N_1 
- g_{1} \bar{N_1} 
(\sigma + i \gamma_5 \vec{\tau} \cdot \vec{\pi}) N_1 
+ \bar{N_2} i \dslash N_2 
- g_{2} \bar{N_2} 
(\sigma + i \gamma_5 \vec{\tau} \cdot \vec{\pi}) N_2 
\nonumber \\
& & 
- \ g_{12} \{ 
\bar{N_1} 
(\gamma_5 \sigma + i \vec{\tau} \cdot \vec{\pi}) N_2 
- 
\bar{N_2} 
(\gamma_5 \sigma + i \vec{\tau} \cdot \vec{\pi}) N_1  
\} 
+ {L}_{\rm mes}  \, , 
   \label{ordsu2lag}
\eeq
where $g_{1}$, $g_{2}$ and $g_{12}$ are free parameters.  
The terms of $g_{1}$ and $g_{2}$ are ordinary chiral invariant 
coupling terms of the linear sigma model.  
The term of $g_{12}$ is the mixing of $N_{1}$ and $N_{2}$.  
Since the two nucleons have opposite parities, $\gamma_{5}$ appears 
in the coupling with $\sigma$, while it does not in the coupling with 
$\pi$.  
The meson lagrangian ${L}_{\rm mes}$ in (\ref{ordsu2lag}) is 
not important in the following discussion.  

Chiral symmetry breaks down spontaneously when the sigma meson 
acquires a finite vacuum expectation value, 
$\sigma_{0} \equiv \bra 0 |\sigma |0 \ket$.  
This generates masses of the nucleons.  
From (\ref{ordsu2lag}), 
the mass can be expressed by a $2 \times 2$ matrix in the space 
of $N_{1}$ and $N_{2}$.   
The mass matrix can be diagonalized by the rotated states, 
\beq
    \left(
    \begin{array}{c}
	N_{+}  \\
	N_{-}
    \end{array}
    \right) 
    = 
    \left(
    \begin{array}{cc}
	\cos \theta      & \gamma_{5} \sin \theta  \\
	- \gamma_{5} \sin \theta &  \cos \theta
    \end{array}
    \right)
    \left(
    \begin{array}{c}
	N_{1}  \\
	N_{2}
    \end{array}
    \right)  \, , \label{eigenN_nai}
\eeq
where the mixing angle and mass eigenvalues are given by 
\beq
\tan 2 \theta = \frac{2g_{1}}{g_{1} + g_{2}} \, , \; \; \; 
m_{\pm} = 
\frac{\sigma_{0}}{2} 
\left(
\sqrt{ (g_{1} + g_{2})^2 + 4g_{12}^2 }
\pm (g_{1} - g_{2} ) 
\right) \, .
\label{anglemass1}
\eeq
In the naive model, since the interaction and mass matrices takes the 
same form, 
the physical states, $N_{+}$ and $N_{-}$, decouple exactly; 
the lagrangian becomes a sum of the $N_{+}$ and $N_{-}$ 
parts.~\footnote{
Small chiral symmetry breaking might induce a small coupling 
$g_{\pi N_{+} N_{-}}$.   
}
Therefore, chiral symmetry imposes no constraint on the relation 
between $N_{+}$ and $N_{-}$.  
The role of chiral symmetry is just the mass generation due to its 
spontaneous breaking. 
When chiral symmetry is restored and
$\sigma_{0} \rightarrow 0$,
both $N_+$ and $N_-$ become massless and degenerate.  
However, the degeneracy is trivial as they are independent; 
they no longer transform among themselves.  
The decoupling of $N_{+}$ and $N_{-}$ implies that the
off-diagonal Yukawa coupling $g_{\pi N_{+}N_{-}}$ vanishes.  
This is a rigorous statement up to the order we considered.

Now we turn to the mirror assignment.  
It is rather straightforward to write down the chiral invariant 
lagrangian compatible to the mirror transformations: 
\begin{eqnarray}
{L}_{\rm{mirror}} & = & 
\bar{N_1} i \dslash N_1 
- g_{1} \bar{N_{1}} 
(\sigma + i \gamma_5 \vec{\tau} \cdot \vec{\pi}) N_{1} 
+
\bar{N_2} i \dslash N_2
- g_{2} \bar{N_{2}} 
(\sigma - i \gamma_5 \vec{\tau} \cdot \vec{\pi}) N_{2} 
\nonumber \\
&-& m_{0}( \bar{N_1} \gamma_{5} N_2 - \bar{N_2} \gamma_{5} N_1  )
+ {L}_{\rm mes} \ .  
    \label{mirsu2lag}
\end{eqnarray}
Here the chiral invariant mass term has been added.  
Note that in the $g_{2}$ term, the sign of the pion field is opposite 
to that of the $g_{1}$ term. 
This compensates the mirror 
transformation of $N_{2}$.  
The lagrangian (\ref{mirsu2lag}) 
was first formulated by DeTar and 
Kunihiro.   

When chiral symmetry is spontaneously broken, the mass matrix of the 
lagrangian (\ref{mirsu2lag}) 
can be diagonalized by a linear combination similar to
(\ref{eigenN_nai}).  
The mixing angle and mass eigenvalues are given by 
\beq
\tan 2 \theta = \frac{2m_{0}}{\sigma_{0}(g_{1} + g_{2})} \, , 
\; \; \; 
m_{\pm} =
\frac{\sigma_{0}}{2}
\left(
\sqrt{ (g_{1} + g_{2})^2  + 4\mu^2 }
\pm (g_{1} - g_{2} )
\right) \, ,
\label{anglemass2}
\eeq
where $\mu = m_{0}/\sigma_{0}$.  
In the mirror model, the interaction term is not diagonalized in the 
physical basis, unlike the naive model.  



Let us show the axial coupling constants $g_{A}$ in the mirror 
model.  
They can be extracted from the commutation relations between 
the axial charge operators $Q_{5}^{a}$ and the 
nucleon fields,
\beq
[Q_{5}^{a}, N_{+}] & = &  \frac{\tau^{a}}{2} \gamma_{5}
(\cos 2 \theta \, N_{+} 
- \sin 2 \theta \, \gamma_{5} N_{-})
\nonumber \\
\ [Q_{5}^{a}, N_{-}] & = & \frac{\tau^{a}}{2} \gamma_{5}
(- \sin 2 \theta \, \gamma_{5} N_{+} - 
  \cos 2 \theta \,  N_{-}) \, . \label{mirq5com} 
\eeq
This implies that $g_{A}$'s are expressed by a 
$2 \times 2$ matrix whose elements are given by the coefficients of 
(\ref{mirq5com}), explaining the result given in (\ref{gAmirror}).  
From this we see that the signs of the diagonal axial charges   
$g_{A}^{++}$ and $g_{A}^{--}$ are opposite.  
The absolute value is, however, smaller than one in contradiction with 
experimental value $g_{A} \sim 1.25$.  
In the present model, the physical states $N_{\pm}$ are the 
superpositions of $N_{1,2}$, whose axial charges are $\pm 1$.
This explains why $|g_{A}^{++}|, |g_{A}^{--}| < 1$.  
In the algebraic method, the $g_{A}$ value can be increased by 
introducing a mixing with higher representations such as 
$(1, \thalf)$.  

\section{$\pi$ and $\eta$ productions at threshold region}

In this section, we propose experimental method to study the two 
chiral assignments.  
As discussed in the preceding sections, one of the differences 
between the naive and mirror assignments is the relative 
sign of the axial coupling constants of the positive and negative 
parity nucleons.  
In the following discussions, we identify $N_{+} \sim N(939)$ and 
$N_{-} \sim N(1535) \equiv N^*$.  
Strictly, the identification of the negative parity nucleon with the 
first excited state $N(1535)$ is no more than an assumption.  
From experimental point of view, however, $N(1535)$ has a 
distinguished feature that it has a strong coupling with an $\eta$ 
meson, which 
can be used as a filter to observe the resonance.  
In practice, we observe the pion couplings which are related to the 
axial couplings through the Goldberger-Treiman relation
$
g_{\pi N_{\pm} N_{\pm}} {f_{\pi}} = g_{A} {M_{\pm}} \, .  
$

\begin{figure}[tbp]
\centering
\epsfxsize = 11cm
\epsfbox{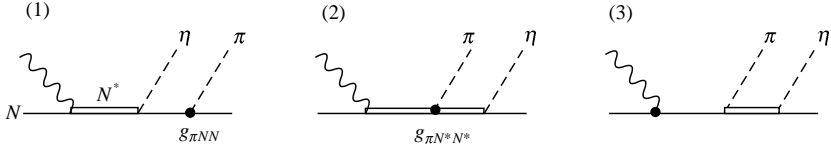}
\begin{minipage}{12cm}
   \caption{
   Dominant diagrams for $\pi$ and $\eta$ productions.  
   The incident wavy line is either a photon or a pion.  
   For photon induced reactions, the diagrams (1) and (2) are 
   dominant, while for pion induced reactions, the third one (3) also 
   contributes.   
   \label{threediag}}
\end{minipage}
\end{figure}

Let us consider $\pi$ and $\eta$ productions induced by
a pion or photon.  
Suppose that the two diagrams of (1) and (2) as 
shown in Fig.~\ref{threediag} are dominant 
in these process.   
Modulo energy denominator, the only difference of these processes is 
due to the coupling constants $g_{\pi NN}$ and $g_{\pi N^* N^*}$.    
Therefore, depending on their relative sign, 
cross sections are either enhanced or suppressed.  
In the pion induced process, due to the p-wave coupling nature, 
another diagram (3) also contributes substantially.

In actual computation, we take the interaction lagrangians:
\beq
L_{\pi NN}
&=& g_{\pi NN} \bar N i \gamma_{5} \vec \tau \cdot \vec \pi N \, ,
\; \; \; 
L_{\eta NN^*}
= g_{\eta NN^*} ( \bar N \eta N^*  + \bar N^* \eta N )  \, ,
\nonumber \\
L_{\pi NN^*}
&=& g_{\pi NN^*} ( \bar N \tau \cdot \pi  N^*
+ \bar N^* \tau \cdot \pi N )  \nonumber \\
L_{\pi N^*N^*}
&=& 
g_{\pi N^* N^*} ( \bar N^* i \gamma_{5} \tau \cdot \pi  N^* )  \, .
\label{Lints}
\eeq
We use these interactions both for the naive and mirror cases with
empirical coupling constants for $g_{\pi NN} \sim 13$, 
$g_{\pi NN^*} \sim 0.7$
and $g_{\eta NN^*} \sim 2$.  
The coupling constants $g_{\pi NN^*} \sim 0.7$
and $g_{\eta NN^*} \sim 2$ are determined from the partial decay widths,
$\Gamma_{N^*(1535) \to \pi N} \approx  \Gamma_{N^*(1535) \to \eta N}
\sim 70$ MeV, although large
uncertainties for the width have been reported.
The unknown parameter is the $g_{\pi N^* N^*}$ coupling.
One can estimate it by using the theoretical value of the
axial charge $g_{A}^{*}$ and the Goldberger-Treiman relation for $N^*$.
When $g_{A}^{*} = \pm 1$ for the naive and mirror assignments,
we find
$
g_{\pi N^* N^*} = g_{A}^{*} m_{N^*}/{f_{\pi}} \sim \pm 17 .
$
Here, just for simplicity, we use the same absolute value as
$g_{\pi NN}$.
The coupling values
used in our computations are summarized in
Table~\ref{parameters}.
\begin{table}[tbp]
    \centering
    \caption{\label{parameters} \small Parameters used in our
calculation. }
    \vspace*{0.5cm}
    \begin{tabular}{ c c c c c c c }
	\hline
	$m_{N}$ & $m_{N^*}$ & $\Gamma_{N^*}$ & $g_{\pi NN}$
	& $g_{\pi NN^*}$ & $g_{\eta NN^*}$ & $g_{\pi N^* N^*}$  \\
	\hline
	938 & 1535  & 140            & 13
	& 0.7     &  2.0    &  13 (naive) \\
	(MeV) & (MeV) & (MeV) &
	&         &         &  --13 (mirror) \\
	\hline
    \end{tabular}
\end{table}

\begin{figure}[tbp]
\centering
\epsfxsize = 12cm
\epsfbox{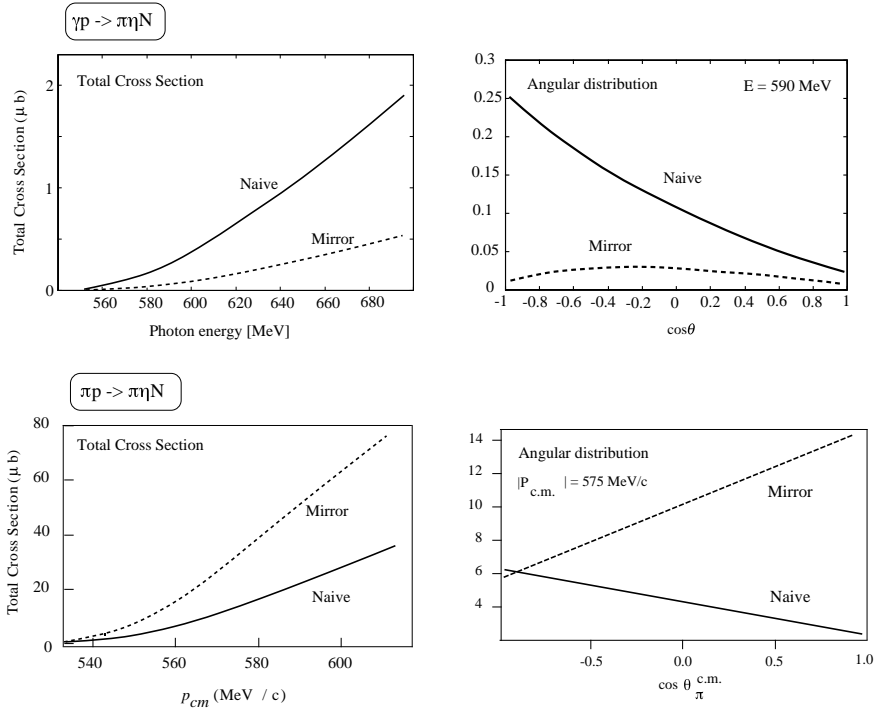}
\begin{minipage}{12cm}
   \caption{
   Various cross sections for $\pi$ and $\eta$ productions.  
   \label{crssct}}
\end{minipage}
\end{figure}

Several remarks follow~\cite{jido2001}:
\begin{itemize}
    \item  We assume resonance ($N^*$)  pole dominance.
    This is considered to be good
    particularly for the $\eta$ production at the threshold
    region, since $\eta$ is dominantly produced by $N^*$.  
    
    \item  There are altogether twelve resonance dominant diagrams.  
    Due to energy denominator, the three diagrams in 
    Fig.~\ref{threediag} are dominant.  
    
    \item  Background contributions, in which two meson (seagull) or 
    three meson vertices appear, are suppressed due to G-parity 
    conservation.  
\end{itemize}
Hence, the processes are indeed dominated by the $N^*$ resonance 
diagrams.  

We show various cross sections for the pion and 
photon induced processes in Fig.~\ref{crssct}.  
We briefly discuss the results:
\begin{enumerate}
    \item  
    The total cross sections are of order of micro barn, which are 
    well accessible by the present experiments.  
    In the photon induced process, the 
    diagrams (1) and (2) interfere destructively in the mirror 
    assignment.  
    In the pion induced case, due to the momentum dependence of the initial 
    vertex the third term (3) becomes dominant and the mirror  
    assignment is rather enhanced.  
    
    
    \item  In the pion induced reaction, the angular distribution of 
    the final state pion differs clearly depending on the sign of the 
    $\pi NN$ and $\pi N^* N^*$ couplings.  
    
\end{enumerate}

\section{Summary}

In this report we have presented an algebraic argument to determine 
the nucleon axial charge $g_{A}$.  
Assuming that the nucleon belongs to a linear representation of the 
chiral group, the commutation relations can determine $g_{A}$.  
This explains the nucleon $g_{A}$ values in the non-relativistic quark 
model, large-$N_{c}$ and the mirror nucleons.  
In order to detect the mirror nucleons, 
we proposed an experiment of $\pi$ and $\eta$ production.   
Various differential cross sections were computed which can be measured 
at the facilities such as SPring8 and LNS, Tohoku.  
Determination of the axial charge is a simple but an interesting 
question related to 
chiral symmetry of the nucleon and
should be studied further both in theory and experiment.  

%
%



\begin{thebibliography}{9}
\bibitem{weinberg1996} 
S. Weinberg, {\it The Quantum Theory of Fields}, 
Cambridge (1996), II p.203.  

\bibitem{weinberg1969} 
S. Weinberg, Phys. Rev. 177 (1969) 2604.  

\bibitem{adler1965}
S. Adler, Phys, Rev. 140 (1965) B736.  

\bibitem{weisberger1966}
W. Weisberger, Phys, Rev. 143 (1966) 1302.  

\bibitem{jido2000}
D. Jido, Y. Nemoto,  M. Oka and A. Hosaka,
Nucl. Phys. {A671} (2000) 471.


\bibitem{jido2001}
D. Jido, M. Oka and A. Hosaka, Prog. Theor. Phys. 106 (2001) 823;\\
D. Jido, M. Oka and A. Hosaka, Prog. Theor. Phys. 106 (2001) 873.

\bibitem{beane}
S.R. Beane, Phys. Rev. D59 (1999) 031901.

\bibitem{manohar}
A.V. Manohar, Nucl. Phys. B248 (1984) 19.

\bibitem{weinberg1990} 
S. Weinberg, Phys. Rev. Lett. 65 (1990) 1177; 1181.  

\bibitem{BWLee} B. W. Lee, {\it Chiral Dynamics}, Gordon and Breach,
New York, (1972)

\bibitem{DeTKun} C. DeTar and T. Kunihiro,
Phys. Rev. D 39 (1989) 2805.

\end{thebibliography}
\end{document}